\documentclass[a4paper]{spie}
\usepackage[dvips]{graphicx}

\title{Mira science with interferometry: a review}

\author{M. Scholz \\
\skiplinehalf
        Institut f{\"u}r Theoretische Astrophysik, Universit{\"a}t
        Heidelberg, Germany, and \\
        School of Physics, University of Sydney, Australia}
\authorinfo{E-mail: scholz@ita.uni-heidelberg.de. Address: Institut f{\"u}r
        Theoretische Astrophysik, Tiergartenstr.15, 69121 Heidelberg, Germany}

\begin{document}
\maketitle

\begin{abstract}
Model-predicted and observed properties of the brightness distribution on
M-type Mira disks are discussed. 
Fundamental issues of limb-darkening and diameter definition, of assigning
observational data to diameter-type quantities and of interpreting such
quantities in terms of model diameters are outlined.
The influence of model properties upon interpretation of measured data is
clarified. 
The dependence of the centre-to-limb variation (CLV) of intensity on 
wavelength, on stellar parameters and on variablity phase and cycle may 
be used for analyzing the geometrical and physical structure of the Mira 
atmosphere, for determining fundamental stellar parameters, and for 
investigating the quality of models. 
Desirable future observations include simultaneous observations in different
spectral features at different phases and cycles, observation of the
position of the shock front and observation of the time- and 
wavelength-dependence of deviations from spherical symmetry.

{\bf The article appeared in SPIE Conf. 4838 (2003), p.163. This version 
contains a reference update including additional references no. 31 and 70.}
\end{abstract}


\section{INTRODUCTION}

Mira science with interferometry began 7 decades ago when Pease (1931)
reported for $o$ Cet a diameter of 47mas measured with Michelson's 
interferometer at Mount Wilson Observatory. A critical assessment of this
result was not published, but some aspects were discussed 37 years later in
Hanbury Brown's (1968) review of stellar diameter measurements. The next step
towards studying Mira variables by means of interferometry was done in a
series of pioneer observations by Labeyrie and co-workers who measured 
diameters of $o$ Cet and R Leo with the speckle interferometry technique
(Bonneau \& Labeyrie 1973; Blazit et al. 1977; Labeyrie et al. 1977; 
Bonneau et al. 1982). From these speckle observations, it became evident that 
diameters measured in different bandpasses of the 0.40 to 1.04$\mu$m region
are so strongly wavelength-dependent that the stellar atmosphere must be
geometrically extended and strong-TiO molecular bands are formed at
substantially larger distances $r$ from the star's center than near-continuum
spectral features. At about the same time, the first high-precision 
measurements of M-type Mira diameters obtained at 
Kitt Peak National Observatory with the 
lunar-occultation technique (U Ori: Ridgway et al. 1977; S Psc, U Ari, 
VV Sgr: Ridgway et al.  1979; S Vir: Ridgway et al. 1980) became available.

The question whether measured differences between monochromatic diameters 
reflect real differences between positions of photon-emitting layers or are 
noticeably affected by differences of the shape of the centre-to-limb variation
(CLV) of intensity was addressed in the model studies of 
Scholz \& Takeda (1987) and Bessell et al. (1989). These authors demonstrated 
(i) that Mira atmospheres are, indeed, geometrically very extended (and must
not be computed in the compact approximation) if dynamic effects taken into 
account; and (ii) that the wavelength-dependence of the CLV shape may 
be strong and has to be accounted for in converting measured visibility
fit diameters into meaningful geometric diameters. Baschek et al. (1991) 
investigated the meaning of various radius-type quantities in case of an
extended atmospheric configuration. Among these, the \emph{intensity radius} is
defined in terms of the CLV shape which, in principle, is an observable 
quantity, whereas the \emph{optical-depth radius} is defined as the distance of
the $\tau$=1 layer ($\tau$ = radial optical depth) from the star's centre, 
which is the usual radius definition in modelling work but no observable 
quantity. 

In the past decade, high-precision interferometric observations and advanced 
models of M-type Mira variables have become available which allow to 
determine physically meaningful diameters and, moreover, may be used for
analysis of the atmospheric structure and for testing the quality of Mira
models. The following topics will be discussed in this contribution:
(a) typical properties of monochromatic CLVs; 
(b) basic problems of defining and measuring monochromatic radii;
(c) concepts of determining "the radius" of a Mira variable to be adopted,
e.g., for describing stellar evolution and pulsation;
(d) monochromatic interferometric observations as a tool for analyzing
the geometrical and physical structure of the Mira atmospheres 
and for scrutinizing modelling assumptions; 
(e) prospects of Mira interferometry including observations of shock front
positions, of deviations from spherical symmetry, and of effects generated
by atmospheric dust in the transition zone from atmospheric to circumstellar 
material. 

The wide field of physics and interferometric observations of circumstellar 
matter and their interpretation will not be reviewed here. Also, this 
contribution does not cover C-type Miras because neither observational nor
theoretical studies of CLV properties and radius interpretation are so far 
available.

\section{MONOCHROMATIC BRIGHTNESS DISTRIBUTION}

Whilst the brightness distribution on compact-atmosphere disks can in most
cases be satisfactorily described in terms of simple CLV approximations 
(see, e.g., Hestroffer 1997; Scholz 1997), among these the uniform disk (UD)
and the fully darkened disk (FDD), such approximations often fail for
Mira variables. CLV reconstructions from lunar occultation data were 
attempted for 4 M-type Miras (Bogdanov \& Cherepashchuk 1991; Di Giacomo et al.
1991) which, despited rather limited accuracy, indicate unusual brightness
profiles. CLV reconstructions from interferometric data are so far not
available, but evidence for strong deviations from conventional moderate
limb-darkening was presented by several authors who reported visibilities 
corresponding to a Gaussian-like brightness distribution (e.g. Wilson et al.
1992; Haniff et al. 1995; Tuthill et al. 1995; Hofmann et al. 2000, 2001;
Young et al. 2000 (S-type Mira $\chi$ Cyg)). Perrin et al. (1999) 
also measured in R Leo a K bandpass visibility that seems to indicate a 
Gaussian-like CLV shape.

The simple exploratory M-type Mira model atmospheres of Bessell et al. (1989) 
predict a large variety of CLV shapes (Scholz \& Takeda 1987), ranging from 
nearly uniform and moderately darkened disks to complex CLVs with, e.g., a
Gaussian-like or a two-component appearance including sometimes small
limb-brightening. More advanced dynamic models and resulting visibilities in 
selected near-continuum to strong-absorption bandpasses confirm this diversity 
(Bessell et al. 1996 (= BSW96); Hofmann et al. 1998 (= HSW98)). From simple 
analytic considerations (Scholz 2001; Mennesson et al. 2002) and model-based 
CLV studies (e.g. HSW98; Jacob et al. 2000; Bedding et al. 2001; 
Jacob \& Scholz 2002 (= JS02)), typical forms of brightness profiles may be 
distinguished in the optical to near-infrared wavelength region: \\
\emph {Small to moderate limb-darkening.} In pure-continuum bandpasses, CLV
shapes are close to the UD (with some modification in the blue due to the
influence of Rayleigh scattering; Scholz \& Takeda 1987; HSW98). Small to 
modest absorption above the continuum-forming layers results in moderate
limb-darkening in many cases. The FDD is a fair approximation and very often
a better approximation than the UD for fitting visibilities. \\
\emph {Gaussian-type CLV.} Strong absorption in substantially cooler layers
than the continuum-forming layers generates a Gaussian-like brightness 
profile in the Wien part of Planck functions. This effect is, in principle, 
independent of atmospheric extension but is more prominent in an extended
atmosphere with high temperature contrasts between high absorbing and deep
continuum-forming layers. \\
\emph {CLV with tail- or protrusion-type extension.} Quite modest 
absorption ($\tau$ noticeably smaller than unity)
in an outer layer of an extended configuration generates a two-component 
CLV appearance in the Rayleigh-Jeans (and near-maximum) part of Planck
functions, consisting of the inner near-UD CLV of the continuum disk
and a tail-type extension of the absorbing outer "shell". Depending upon
the geometric  dimensions of this extension, visibilities may be moderately
to strongly affected and even "distorted". As absorption increases and
absorbing layers move closer to continuum-forming layers, tails resemble 
more protrusions of the inner CLV. The Gaussian-like CLV appearance found 
by Perrin et al. (1999) probably is produced by this kind of protrusion due 
to H$_2$O contamination of the K bandpass continuum (JS02). 
Details depend on the geometric position and on the geometric and 
optical thickness of the absorbing "shell" (cf. Tej et al. 2003). \\
\emph {Uncommon CLV.} Uncommon brightness distributions that cannot be
described in terms of a simple scheme and that are not found in conventional 
compact-atmosphere stars may occur, depending on wavelengths, for intricate
atmospheric stratifications.

Clearly, complex shapes of the monochromatic brightness distribution
require accurate multi-baseline interferometric data that cover a large 
range of spatial frequencies if sound information is to be 
extracted from observations. For instance, naive visibility fitting 
procedures may lead to physically meaningless diameter-type quantities 
(e.g. HSW98; Bedding et al. 2001; JS02).

\section{MONOCHROMATIC INTENSITY DIAMETER AND FIT DIAMETER}

The CLV is the only geometric information about the stellar disk the 
telescope receives. A diameter-type quantity that is defined in terms of
the shape of the CLV is called an intensity diameter (Baschek et al. 1991).
In the case of the optical solar disk, the position of the CLV inflection
point is used for defining the photospheric diameter of the Sun (see, e.g.,
Minnaert 1953). This diameter is a wavelength-dependent quantity but $\lambda$
variations are minute in the optical wavelength region. Essentially, the
inflection point marks the position of the steep CLV flank at the edge of 
a compact-atmosphere stellar disk with small to moderate limb-darkening.
In principle, an intensity radius defined in terms of the position of this
flank may be applied to stars when high-precision CLV reconstructions become 
available.

For Mira disks, pure-continuum brightness profiles have steep flanks which
accurately mark the position of continuum-forming layers (e.g. JS02). Weak 
molecular-band contamination of the continuum usually results in a slightly 
more limb-darkened CLV which still has a steep flank suited for defining a
physically meaningful intensity diameter, but other types of CLV shapes do 
also occur depending on wavelength and on the position and thickness of the 
absorbing layers. In particular, tail- or protrusion-type extensions of the
continuum CLV lead to multiple inflection points and may even involve slight 
limb-brightening (cf. Scholz 2001; Mennesson et al. 2002; JS02). No recipe 
is known for predicting a priori the influence of weak contamination 
upon the brightness profile. Rather, a case-by-case investigation has 
to be performed.

In the case of Gaussian-type CLVs, the FWHM may be adopted as radius-type
quantity describing the disk's brightness profile. There exists, however, no 
straight-forward correlation between the FWHM or the position of the inflection
point and the position of, e.g., the $\tau$=1 layer in the absorbing 
spectral feature that generates this Gaussian appearance (e.g. Scholz 
1997, 2001; HSW98; Jacob et al. 2000).

We conclude that the definition of meaningful intensity diameters in terms 
of the shape of the intensity CLV may be very difficult or
almost impossible in the light of many spectral features, even at 
near-continuum wavelengths. Extremely high accuracy of CLV reconstruction
from interferometry would be required in order to determine whether and where
meaningful radius points on the CLV do exist.

In principle, fit diameters which are obtained by fitting an observed
visibility by the visibility of an artificial brightness distribution 
(UD, FDD, Gauss, ...) are a variant of intensity diameters because they
define a diameter-type quantity in terms of the CLV shape. In the case
of a simple brightness profile with small to moderate limb-darkening,
both UD and FDD fits provide a good approximation of the position of the
CLV flank. The model study of HSW98 shows that they typically do not 
deviate from each other by more than about 10 percent and that FDD fits 
tend to match the real CLV significantly better. It also shows that
visibilities from Gaussian-like brightness profiles are very poorly 
approximated by UD or FDD fits and may readily be recognized by 
considering a wide range of spatial frequencies including the 
neighbourhood of the first null. Single-baseline interferometric
observations may lead to physically meaningless UD or FDD fit diameters 
depending on the spatial frequency of the fit. Unfortunately, a Gauss fit 
only yields the width (e.g. FWHM) of the brightness decline but no direct 
information about the geometric dimensions of the stellar disk. Similar 
sytematic studies of observed visibilities have yet to be performed, but the
occurrence of Gaussian-like CLVs in certain bandpasses is well
established (see references listed in Section 2).

Particular problems occur in case of a two-component appearance of the
brightness distribution generated in the Rayleigh-Jeans (or near-maximum)
part of Planck functions by modest molecular absorption in upper
atmospheric layers (see Section 2). This sitation is predicted by models in 
standard near-continuum bandpasses in the near-IR (e.g. H, K, L) and may lead 
to distorted visibilities, very poor UD fits and a strong baseline-dependence
of fit diameters in case of single-baseline measurements. The relevance and
interpretation of such diameters was investigated in the model study of JS02. 
High-precision interferometry should be able to identify such visiblity forms
and to extract information about the position of both the continuum-forming 
and the upper absorbing layers (JS02; Tej et al. 2003).

\section{MONOCHROMATIC OPTICAL-DEPTH DIAMETER}

The monochromatic optical-depth radius $R_{\lambda}$ is the distance of the 
$\tau_{\lambda}$=1 layer from the star's centre. This radius definition is a 
standard definition in stellar models but no observable quantity. Since the
radial optical-depth interval d$\tau_{\lambda}$ = 
-$k_{\lambda}$($r$)$\rho$($r$)d$r$ = -d$r$/$l_{\lambda}$($r$) ($\rho$ =
density; $k_{\lambda}$ = extinction coefficient) measures the local distance
interval d$r$ in units of the local photon mean free path $l_{\lambda}$($r$),
this choice of radius definition means choosing the layer which is just one
integrated mean free path below the "surface" of the stellar atmosphere. 
Unfortunately, however, photons collected by the observer 
often originate from a wide range of depths around
the $\tau_{\lambda}$=1 layer, and there is no trivial correlation between the
shape of the CLV curve and the position of $R_{\lambda}$ on that curve (e.g.
Scholz 1997, 2001; Jacob et al. 2000; JS02).

Since optical-depth diameters are model quantities, they depend in principle
not only upon the global stellar parameters (see Section 6) including phase
and cycle of variability but also upon the quality of the model. We refer to
the original papers regarding modelling approximations (Bessell et al. 1989;
BSW96; HSW98;
H\"{o}fner et al. 1998; Woitke et al. 1999)
and only list the most critical points: 
(a) drive of pulsation (e.g. piston BSW96; self-excited HSW98);
(b) dynamical density stratification resulting from shock-front-driven 
outflow and subsequent infall of matter; 
(c) non-grey temperature stratification which, in presently available models, 
is computed in the approximations of local thermodynamic
and radiative equilibrium (cf. Beach et al. 1988 and Bessell et al. 1989 
as for relaxation of shock-front-heated material); 
(d) completeness and numerical treatment of molecular band absorption,
including serious and unsolved problems of applying opacity distribution 
function techniques (cf. Baschek et al. 2001; Wehrse 2002) and 
opacity sampling techniques, respectively, to a 
dynamic atmosphere with pronounced velocity stratification; 
(e) treatment of the transition zone from atmospheric to circumstellar 
layers (including dust formation).

Since intensity diameters are not accessible to presently available 
interferometric observations, and since the physical relevance of such
intensity diameters is quite modest in many cases, measured fit
diameters are usually converted into optical-depth diameters via
model considerations. This requires, however, that models with suitable
stellar parameters (including variability phase and cycle) 
are chosen and that sufficiently realistic models
are available. Otherwise, derived optical-depth diameters are more or less
formal quantities without sound correlation to the geometric structure of the
Mira atmosphere. Generally, the conversion of a fit diameter into an
optical-depth diameter tends to be safest in cases of steep-flank CLVs with
$R_{\lambda}$ lying on the steep flank. In particular, real-continuum
radii $R_{\lambda {\rm cont}}$ are perfectly marked by the flank position
(e.g. JS02).

\section{FILTER-INTEGRATED QUANTITIES}

Real observations are done in bandpasses of finite width, i.e. a physically
meaningful interpretation of secured data requires accurate knowledge of
filter transmission $f_{\lambda}$ unless the CLV is $\lambda$-independent 
(e.g. pure continuum) inside the considered bandpass. No trivial correlation 
between the monochromatic CLVs inside a filter and the filter-integrated CLV 
exists. Obviously, high-intensity portions of the spectrum contribute more
strongly to the CLV than low-intensity portions. This means that the integrated
CLV tends to be dominated by contributions of deeper layers belonging to
smaller monochromatic diameters.

The optical-depth radius of a bandpass may either be defined in terms of some
mean value of monochromatic optical-depth radii $R_{\lambda}$ or in terms of
a mean optical depth belonging to a mean extinction coefficient. The most 
common definition of the filter radius by Scholz \& Takeda (1987) uses the
central intensity on the disk $I_{\lambda}^{\rm c}$ as weighting function of
$R_{\lambda}$:
$R_{\rm f} = \int R_{\lambda} I_{\lambda}^{\rm c} f_{\lambda} {\rm d}{\lambda}
             / \int I_{\lambda}^{\rm c} f_{\lambda} {\rm d}{\lambda}$ . 
For safe conversion of measured intensity or fit radii into optical-depth 
radii, the filter transmission has to be known accurately. Bandpass radii
based on mean optical depths are uncommon but are being used for defining
"the radius" of the star averaged over all wavelengths (see Section 6).

Clearly, impure filters that contain a mixture of low-intensity and 
high-intensity spectral features are hardly useful for extracting information
about the geometry and physics of the Mira atmosphere. Special care has to be
taken in the choice of bandpasses that are to probe strong-absorption 
features of the spectrum. If these bandpasses have leaking transmission
wings and intensities are substantially higher in the wings, the observation
may be dominated by the weak-absorption features although the bandpass is
centred in the strong-absorption feature and only contributions from this 
feature are within the FWHM bandwidth. This may happen, e.g., in the Wien
part of Planck functions where the extreme temperature sensitivity of
the source function results in extreme intensity contrasts between strong-
and weak-absorption features.

Note in this context that picket-fence structures of certain molecular 
bands in which near-continuum gaps appear between strong lines also produce 
impure-filter-like effects, and the line spectrum of suspect molecular bands 
(e.g. CO) in the bandpass should be checked before starting an interferometric 
project.

\section{MIRA AND PARENT STAR DIAMETER}

The fundamental parameters of a variable star are the mass $M$, the 
time-dependent luminosity $L(t)$ and the time-dependent radius $R(t)$. In
the case of an extended-atmosphere star such as a Mira variable, a
sensible definition of "the radius" of the star has to be agreed upon. The
most common definition refers to the position of the atmospheric layer where
the radial Rosseland optical depth equals unity, $R$=$r(\tau_{\rm Ross}=1)$, 
and this definition is called the Rosseland radius. The effective temperature 
$T_{\rm eff} \propto (L/R^2)^{1/4}$ of the star always refers to a specific
radius definition.

Several caveats have to be noted. (a) The Rosseland radius is an optical-depth
radius, i.e. it is a model quantity but no observable quantity. (b) The
Rosseland opacity is a harmonic-type mean opacity weighting heavily the
low-extinction portions of the spectrum. Its calculation may depend strongly 
on the completeness and numerical treatment of molecular band absorption (see
Section 4). For instance, Rosseland opacities used for constructing interior 
models are often based on substantially more simplified 
molecular band contributions than
those used in atmospheric models. (c) In most cases, the heavy weight of
low-extinction contributions to the Rosseland average means that $R$ is close
to the continuum radius $R_{\lambda {\rm cont}}(t)$ which depends little on
$\lambda$ (see Scholz \& Takeda 1987 and HSW98 for effects of Rayleigh 
scattering in the blue) and is in principle observable by determining
the position of the steep CLV flank (see Section 4). (d) At very cool phases,
molecular absorption may blanket near-continuum windows so efficiently that
the Rosseland mean increases significantly and the Rosseland radius is
shifted into much higher layers than the continuum-forming layers (e.g. HSW98).

For these reasons, a sensible alternative definition of "the radius" in terms
of a continuum radius may be prefered. Standard near-continuum bandpasses
in the near-IR would be suited if slight to moderate molecular contamination 
is carefully accounted for (cf. JS02).

The above discussed radii are time-dependent quantities. If time-independent 
properties of Mira variables are investigated, such as the position and (slow)
evolution in the HR-diagram, the period-luminosity-relation or the 
period-radius-relation, time-independent luminosities and radii 
have to be used. Whilst the time-independent luminosity $L_{\rm p}$ is the 
time-average of $L(t)$, there exists no physically meaningful time-average
of $R(t)$. Instead, the "parent star" of the Mira variable is considered. This
is a hypothetical non-pulsating star with the same mass $M$ and 
luminosity $L_{\rm p}$ as the Mira variable and with the Rosseland radius 
$R_{\rm p}$. Models predict (BSW96; HSW98) that $R(t)$ may
typically deviate from $R_{\rm p}$ by up to about 10 or 20 percent, but
larger deviations may occur at very low effective temperatures owing to the
above mentioned behaviour of Rosseland opacities.

Note that the parent star is an artificial object, and deriving $R_{\rm p}$ 
or any other radius of the parent star from observations is possible only
via model considerations.

A wealth of interferometric measurements of "the radius" of M-type Mira
variables has been reported in recent years (e.g.
Karovska et al. 1991; 
Haniff et al. 1992, 1995; 
Quirrenbach et al. 1992;
Ridgway et al. 1992;
Wilson et al. 1992;
Tuthill et al. 1994, 1995, 1999, 2000;
van Belle et al. 1996, 1997 (S-type Miras), 2002;
Weigelt et al. 1996;
Bedding et al. 1997; 
Lattanzi et al. 1997 (HST observations); 
Burns at al. 1998; 
Perrin et al. 1999; 
Hofmann et al. 2000, 2001, 2002; 
Weiner et al. 2000 (mid-IR);
Young et al. 2000 ($\chi$ Cyg);
Mennesson et al. 2002;
Thompson et al. 2002a, 2002b).
The degree of elaboration of reduction of the measured fit diameter 
to a physically meaningful geometric size parameter varies greatly. Quite 
often, it is assumed that a UD-fit radius in a near-continuum spectral bandpass
is close to $R_{\lambda {\rm cont}}(t)$ as well as to the Rosseland radius 
$R(t)$ and, when the position in the HR-diagram or the period-radius-relation
is considered, that $R_{\lambda {\rm cont}}(t)$ or $R(t)$ 
is a fair approximation to $R_{\rm p}$. Comparison of measured diameters and 
deduced effective temperatures from different sources has to account for
the radius definition adopted by the observer. The fundamental problems of 
converting a Mira fit diameter observed in near-continuum bandpasses into 
a physically meaningful quantity and strategies of minimizing uncertainties
are discussed in JS02.

The period-radius-relation was investigated by several authors
(e.g. Haniff et al. 1995; Van Belle et al. 1996; Van Leeuwen et al. 1997;
Whitelock and Feast 2000)
and seems to point to first-overtone-mode pulsation of M-type Mira variables
(cf., however, a different interpretation given by Ya'ari \& Tuchman 1999). 
No strict reduction of measured data to the parent star radius was carried out
in these studies. More recent work (e.g. Perrin et al. 1999; Hofmann et al.
2002; Mennesson et al. 2002; van Belle et al. 2002) and the model study of 
JS02 suggest sytematically smaller radii to be inserted in the 
period-radius-relation and, hence, rather favours fundamental-mode pulsation
in agreement with other pulsation-mode indicators (see, e.g., Wood et al. 1999;
Scholz \& Wood 2000).

Observations of the pulsation of a Mira, i.e. of the variation of the
continuum diameter with phase were reported by, e.g., Tuthill et al. 
(1995, $o$ Cet), Burns et al. (1998, R Leo) 
, Perrin et al. (1999, R Leo)
and Young et al. 
(2000, $\chi$ Cyg), but the problem of the influence of phase-dependent 
molecular contamination upon measured fit diameters is still not fully 
solved (JS02).

\section{MODEL DEPENDENCE OF DIAMETER MEASUREMENTS}

The only way of securing through interferometry 
direct information about the geometry and physics of a Mira 
would be the full reconstruction of the monochromatic or
filter-integrated CLV at different wavelengths. These CLVs could be compared
to model predictions. Such CLV reconstructions are not feasible in the
nearest future. In practice, observers obtain fit diameters at different levels 
of sophistication and accuracy and have to convert these fit diameters into 
more meaningful geometric quantities via model considerations.

In the ideal case, a "perfect model" should be available that has the same
fundamental stellar parameters ($M$, $L(t)$, $R(t)$) and the same variablity 
phase as the observed star. In fact, as Miras show more or less pronounced 
cycle-to-cycle-dependence of the atmospheric structure, a set of such
models covering a large number of cycles would be required. When neither the
fundamental parameters of the observed star nor the characteristics of the 
cycle of observation are already known, the conversion of fit diameters into 
physically more meaningful diameters would have to be an iterative process.

At the present stage of Mira model construction, such models are far from being
perfect (see Section 4; BSW96; HSW98; H\"{o}fner et al. 1998; Woitke et al.
1999), are available for only a very few 
stellar parameters, and cover only a few selected phases of a small number of 
successive cycles. Thus, interpretation of measurements and deduced 
"observed diameters" may be model-specific and may change when different 
models are used.

\section{MIRA ANALYSIS BY INTERFEROMETRY}

High-accuracy interferometric observations may not only be used to determine
"the radius" of the Mira star (see Section 6) but may also serve as a tool of 
diagnostics of the geometric and physical structure of the Mira atmosphere.
Since the atmospheric stratification depends on all fundamental stellar
parameters ($M$, $L$, $R$), precise interferometry may also be used to obtain
information about the otherwise unobservable stellar mass. 

As the interpretation of measured data as well as the analysis of the 
atmospheric stratification are model-dependent, and as the correlation 
between this stratification and the stellar parameters has to be derived 
from a model, results are coupled to the availablity of reliable models in 
the relevant ranges of stellar parameters and phase-cycle-combinations.
In addition, high-accuracy interferometry in combination with spectroscopy
(cf. Tej et al. 2003) may be used to investigate the quality and possible
shortcomings of Mira modelling approximations (see Section 4).

The most promising approach to interferometric Mira analysis consists of
(i) simultaneous observations with different baselines (in order to explore
the shape of the CLV), (ii) simultaneous observations in different bandpasses
(in order to probe the momentaneous atmospheric stratification), and (iii) 
observations in the same bandpasses at different phase-cycle-combinations
(in order to probe the time variation of the atmospheric stratification).
Observations at the same phase of different cycles may be combined although
cycle-independence may not be taken as granted and has to be checked in
each case. Numerous observations in different bandpasses (though not
alway secured at the same time) have been reported in the past decade (e.g.
Haniff et al. 1992, 1995;
Tuthill et al. 1995, 1999, 2000;
Weigelt et al. 1996;
Burns et al. 1998;
Hofmann et al. 2000, 2001;
Young et al. 2000 ($\chi$ Cyg);
Mennesson et al. 2002;
Thompson et al. 2002b).
No sytematic attempts have so far been made to use such observations for
analyzing the atmospheric structure. A very promising spectroimaging
technique which provides direct access to molecular contributions shaping the
disk brightness was successfully applied to $o$ Cet (Takami et al. 2003).
Model-based studies that discuss wavelength effects include, e.g., HSW98, 
Hofmann et al. (2000, 2001), Jacob et al. (2000), and JS02. 

One of the most urgent and most rewarding observations for probing the
Mira atmosphere would be the determination of the position of the 
shock front. So far, only one observational project was designed for this
purpose and was not successful for technical reasons (P.~R. Wood, private 
communication, contact wood@mso.anu.edu.au for information).
The hot post-shock material is seen in the spectrum by
typical emission lines, in particular the Balmer series of hydrogen. The
hot zone is very narrow (Fox \& Wood 1985), and its influence on atmospheric 
temperatures and photon fluxes is usually neglected in Mira models (BSW96; 
HSW98; see also Beach et al. 1988 and Bessell et al. 1989). 
Balmer-line fluxes are, however, strong enough to be accessible to
high-precision interferometry (Fox et al. 1984; Gillet et al. 1983, 1985a,
1985b; Gillet 1988a, 1988b; Woodsworth 1995 (S-type Miras); Richter \& Wood
2001). 
Line strengths vary strongly with phase (and cycle). In principle, any Balmer 
emission line may serve the purpose of measuring the shock front position, 
though H$\alpha$ may be less suited because of much higher fluxes from 
the cool atmospheric material at this wavelength.

Another unsolved problem is the frequency, time-dependence and 
wavelength-dependence of observed deviations from circular symmetry of the 
Mira disk. Tentative explanations include, e.g., surface spots, 
non-radial pulsation modes and asymmetric shock fronts, patchy high-layer
extinction, effects of a circumstellar dust or gas disk, rotation, and the 
influence of a companion star. Numerous observations of asymmetries have been 
reported in the literature (e.g. 
$o$ Cet: Karovska et al. 1991, Haniff et al. 1992, Wilson et al. 1992,
   Quirrenbach et al. 1992, Haniff et al. 1995, Lattanzi et al. 1997 (HST), 
   Tuthill et al. 1999; 
$\chi$ Cyg: Haniff et al. 1995, Tuthill et al. 1999;
R Cas: Haniff et al. 1995, Weigelt et al. 1996, Tuthill et al. 1999, 
   Hofmann et al. 2000; 
R Dor: Bedding et al. 1997;
W Hya: Lattanzi et al. 1997, Tuthill et al. 1999;
R Leo: Tuthill et al. 1999;
R Tri: Thompson et al. 2002a),
but little is known about phase-cycle-dependence or pulsation-independent 
time variation in given bandpasses and about wavelength-dependence at a given
time. Systematic studies including closure phase measurements would not only 
reveal such dependences but would also allow to disentangle contributions to
two-component CLV shapes generated by spherical molecular "shell" absorption
(see Section 2) and by surface spots.

This review does not cover interferometry of the circumstellar environment of
Mira stars. Still, one has to be aware that there is no real separation of
atmospheric and circumstellar layers. Model makers use to choose the "surface"
of the stellar atmosphere in terms of an estimated very small 
optical depth which is so small that both radiative energy transport 
and strongest spectral absorption
features can be treated correctly. In the BSW96 and HSW98 models, the distance
of this "surface" from the star's centre ranges from 2$R_{\rm p}$ to 
5$R_{\rm p}$ for different model series. In particular, the occurrence of
dust in upper atmospheric layers or in the stellar-to-circumstellar transition
layers would considerably affect brightness distributions (cf. Bedding et al.
2001). Interpretation of interferometric and spectroscopic data indicates,
indeed, that the inner edges of dust shells might be as close as 2 or 3 
continuum radii from the star's centre (e.g. Danchi et al. 1994; 
Danchi \& Bester 1995; Lobel et al. 2000; Lorenz-Martins \& Pompaia 2000).
Although the uncertainty of such values is hard to assess, it is evident from
the exploratory model study of Bedding et al. (2001) that dust may form 
under certain conditions in upper atmospheric layers and would show 
significant effects on brightness distributions and spectral 
characteristics in the optical to near-IR wavelength range with typical 
phase-cycle- and wavelength-dependence.

\section*{ACKNOWLEDGEMENTS}

I wish to thank my colleagues from interferometry groups all over the world
for many fruitful discussions.

\end{document}